# Triple V-shaped type-II quantum wells for long wavelength interband cascade lasers

B. Petrović,[1] T. Sato,[2,3] S. Birner,[2] J. Zanon[4], M. E. Flatté[5,4], R. Weih,[6] F. Hartmann,[1] and S. Höfling[1]

[1]*Julius-Maximilians-Universität Würzburg, Physikalisches Institut, Lehrstuhl für Technische Physik, Am Hubland, 97074 Würzburg, Germany*

[2]*nextnano GmbH, Konrad-Zuse-Platz 8, 81829 München, Germany*

[3]*TUM School of Computation, Information and Technology, Technical University of Munich, Hans-Piloty-Straße 1, 85748 Garching, Germany*

[4]*Technische Universiteit Eindhoven, Department of Applied Physics, Groene Loper 19, 5612 AP Eindhoven, Netherlands*

[5]*Department of Physics and Astronomy, University of Iowa, Iowa City, Iowa 52242, USA*

[6]*nanoplus Advanced Photonics Gerbrunn GmbH, Oberer Kirschberg 4, 97218 Gerbrunn, Germany*

We investigate triple V-shaped type-II quantum wells designed to emit at 6-9 μm wavelength range, consisting of three InAs(Sb) electron quantum wells and two $Ga_{0.6}In_{0.4}Sb$ hole quantum wells. The wells' composition and thicknesses are optimized in terms of wavefunction overlap and valence intersubband absorption (VISA) by using $\mathbf{k}\cdot\mathbf{p}$. calculation. The triple V-shaped type-II quantum well designed for emission at 6.2 μm, 7.5 μm and 9.0 μm, show 18.8 %, 18.9 % and 19.2 % higher wavefunction overlap respectively, compared to the corresponding W-shaped designs of $InAs/Ga_{0.6}In_{0.4}Sb/InAs$ wells commonly employed in active regions of interband cascade lasers. In addition, the calculated VISA in V-shaped quantum wells for the designed wavelength is significantly smaller than in W-shaped wells. These enhancements might extend the wavelength limit of room temperature emission of GaSb-based interband cascade lasers.

## 1. Introduction

Interband cascade lasers, (ICLs [1]) are one of the commonly used coherent light sources in the mid-infrared wavelength region. The active region design of ICL combines interband recombination mechanism of the laser diodes with cascading structure introduced for quantum cascade lasers [2]. Multi-photon generation for a single electron-hole pair, along with broad spectral tunability provided by type-II band alignment, enables low threshold current densities in the range 3-7 μm at room temperature (RT) [3]. This spectral window is of interest for various gas-sensing applications such as environmental monitoring [4], industrial control [5] and exhaled breath analysis for medical diagnosis [6]. The first RT continuous wave (cw) operation of ICLs has been achieved at wavelength of 3.75 μm in 2008 [7]. At the "sweet spot" wavelength range of 3.6-3.7 μm the threshold current density of broad area devices at RT has been significantly decreased (134 A/cm$^2$) [8] by carrier rebalancing [9]. As shown in Ref. 9, the ICLs performance can be enhanced by heavily doping the electron injectors, which compensates for the excess hole density in the recombination region. The lowest reported threshold current density for ICLs (98 A/cm$^2$) has been obtained by increasing the number of cascade stages [10].

At longer wavelengths, achieving lasing emission at RT becomes more difficult since free carrier absorption increases, whereas due to the wider quantum wells the optical matrix element decreases. As an additional important figure of merit, valence intersubband absorption (VISA) has been evaluated from the eight-band $\mathbf{k}\cdot\mathbf{p}$ model of the active region of the ICL, at the wavelength range 2-10 μm [11-12]. A recombination region of an ICL consists of type-II "W-shaped" quantum wells [13] (QWs), where electrons are dominantly confined in InAs wells while the hole-confining GaInSb-well is sandwiched between them. It was shown that a careful design of the hole GaInSb QW mitigates VISA losses. Specifically, the width of GaInSb well corresponding to the lowest absorption is 2.5 nm, for wavelengths shorter than 4.8 μm, 3.0 nm for the range 4.8-5.4 μm and 3.5 nm for wavelengths longer than 5.4 μm. Design optimization has been demonstrated experimentally as for the ICL at 4.3 μm close to record low threshold current density was obtained [11]. Following the same model, ICL designed to emit at 6.2 μm has shown a RT emission in cw operation at the longest wavelength reported so far [14]. Conversely, in pulsed mode, the longest wavelength at which GaSb-based ICL has been operated at RT is 6.8 µm [15]. The limit for InAs-based ICLs of 7.0 µm [16] has most recently been extended to 7.7 µm [17] by implementing two-stage InAsP-AlAsSb barriers in the recombination region, similar to what was done in earlier reports for longer wavelengths [18-19]. One of the major obstacles for achieving RT emission is the small oscillator strength and consequently the wavefunction overlap in the recombination region. Moving towards longer wavelengths, the emission of InAs-based ICLs has been demonstrated at cryogenic temperatures by using heavily doped plasmon-enhanced claddings [18-20]. In addition, recent progress has been made on Si-based ICLs in achieving comparable performance to state-of-the-art GaSb-based ICLs at 3-4 µm emission sweet spot [21-22]. However, in the wavelength range 7-8 µm, for Si-based lasers, RT operation has only been realized for QCLs [23-24].

The wavelength of 7.5 µm is of a particular interest as it is range of absorption lines of conglomerate of molecules, such as TNT (7.3 µm), $SO_2$ (7.2 µm), $H_2S_2$ (7.3 µm), $H_2O_2$ (7.8 µm) and $H_2SO_4$ (7.9 µm). Covering the absorption line of TNT is crucial for explosive detection, while sulfur-based oxides and acids in gaseous form are colorless, odorless and toxic for inhaling, therefore their suppression is of great importance for health safety. Additional neighboring wavelengths of interest are 6.2 µm, the absorption line of $NO_2$ and 9.0 µm, the absorption line of glucose and ethanol.

In this work, a V-shaped QW design consisting of **AlSb**/*InAs*/$Ga_{0.6}In_{0.4}$Sb/*InAs$_x$Sb$_{1-x}$*/$Ga_{0.6}In_{0.4}$Sb/*InAs*/**AlSb** is compared to the standard W-shaped QW (**AlSb**/*InAs*/$Ga_{0.6}In_{0.4}$Sb/*InAs*/**AlSb**) in terms of wavefunction overlap and VISA for wavelengths 6.2 µm, 7.5 µm and 9.0 µm. In the second section, layer structure optimization for maximizing wavefunction overlap is carried out for both W-shaped and V-shaped design. Specifically, for 7.5 µm design, wavefunction overlap and VISA are investigated for applied electric field in the third section.

**2. Optimization of V-QW at zero electric field**

The aim of this work is to enhance the design of the recombination region of the 7.5 µm ICLs in terms of the wavefunction overlap and VISA. Such attempts have been made by introducing multiple InAs QWs to the original W-shaped recombination region of interband cascade lasers [25-26] for emission wavelengths of 4.5 µm and 7 µm, [27] ~5.4 µm, [28] and ~4.0 µm [29]. We consider a similar design of triple V-shaped QWs consisting of **AlSb**/*InAs*/$Ga_{0.6}In_{0.4}$Sb/*InAs$_x$Sb$_{1-x}$*/$Ga_{0.6}In_{0.4}$Sb/*InAs*/**AlSb**. In this design, two GaInSb-wells are employed instead of one. They are sandwiched between InAs wells from outer and InAsSb well from inner side. A fraction of antimony is used to improve the electron confinement in the middle QW. To provide a clear comparison between W-shaped and triple V-shaped QWs, we are optimizing design parameters of both structures to maximize the wavefunction overlap. Firstly, and for simplicity, the wells will be considered without an electric field applied. The band edge diagram of the optimized designs are presented in Figure 1 (a) for W-shaped (design A) and Figure 1 (b) for V-shaped QW (design B). We obtain the zone-center carrier wavefunctions and subband energies in these type-II QWs under strain by solving the effective-mass Schrödinger equation by the finite difference method implemented in the software nextnano[3] [30]. All the material parameters were taken from [31]. In the W-shaped QW of the layer sequence **AlSb**/*InAs*/$Ga_{0.6}In_{0.4}$Sb/*InAs*/**AlSb**, the maximum wavefunction overlap in the wavelength range 7.4-7.6 µm was achieved for widths (**2.50**/*2.65*/2.30/*2.65*/**1.50**) nm. Similarly, the maximal wavefunction overlap in the wavelength range 6.1-6.3 µm was obtained for the layer sequence (**2.50**/*2.40*/2.20/*2.40*/**1.50**) nm, whereas in the wavelength range 8.9-9.1 µm, the maximum wavefunction overlap was obtained for design (**2.50**/*2.95*/2.20/*2.95*/**1.50**) nm.

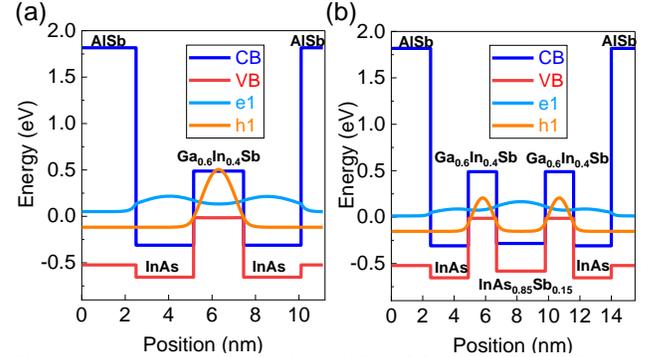

Fig. 1. (a) Band edges of the W-shaped QW at 7.5 µm wavelength: Design A: **AlSb**/*InAs*/$Ga_{0.6}In_{0.4}$Sb/*InAs*/**AlSb**, (**2.50**/*2.65*/2.30/*2.65*/**1.50**) nm

(b) Band edges of the triple V-shaped QW at 7.5 µm wavelength: Design B: **AlSb**/*InAs*/$Ga_{0.6}In_{0.4}$Sb/*InAs$_{0.85}$Sb$_{0.15}$*/$Ga_{0.6}In_{0.4}$Sb/*InAs*/**AlSb**, (**2.5**/*2.4*/1.8/*3.1*/1.8/*2.4*/**1.5**) nm.

Figure 2 (a) shows the wavefunction overlap as a function of the widths of electron InAs and hole $Ga_{0.6}In_{0.4}$Sb QW. At wavelengths of approximately 7.5 µm, the wavefunction overlap varies from 0.516 to the maximum value of 0.540 (design A). Designs of the 7.5 µm wavelength are plotted in blue contour. Designs with shorter (longer) wavelengths belong to the area on the left- (right-) hand side, respectively. At longer wavelengths, the wavefunction overlap values are below 0.516, while at shorter wavelengths the wavefunction overlap is above 0.540. Along with 7.5 µm, designs of wavelengths 6.2 µm and 9.0 µm are marked with green and purple contour. Maxima of wavefunction overlaps at 6.2 µm and 9.0 µm are 0.573 and 0.510, respectively.

Figure 2 (b) depicts a horizontal cut of Figure 2 (a) across Design A, which is marked by dots. The figure shows that the wavelength increases with thickness of InAs QWs. Similarly, Figure 2 (c) presents a vertical cut of Figure 2 (a) across Design A. The wavelength is increasing parabolically with the thickness of $Ga_{0.6}In_{0.4}Sb$ QW. The maximum of about 8.15 μm is reached at 3.5 nm wide QW, above which the energy of the hole eigenstate does not change further with increasing the well width. The increase of the wavelength, however, is not as prominent as in the previous case due to larger effective mass of holes in $Ga_{0.6}In_{0.4}Sb$. For both cuts, we note that the wavefunction overlap is almost linearly decreasing with well width. Similarly to wavelengths, the wavefunction overlap is more sensitive to the thickness of InAs QWs.

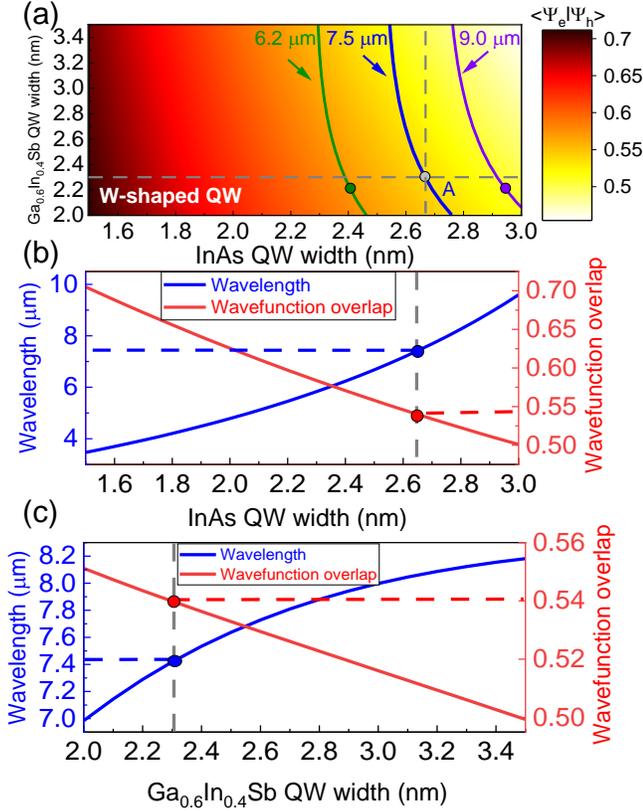

Fig. 2. (a) Dependence of the wavefunction overlap in a W-shaped QW for a variation of the electron (InAs) and the hole ($Ga_{0.6}In_{0.4}Sb$) quantum well (QW) widths. The region corresponding to wavelength of 7.5 μm is marked with the solid, blue contour line. Design A, which corresponds to the highest wavefunction overlap, is highlighted by the gray dot. The regions corresponding to wavelengths of 6.2 μm and 9.0 μm are marked with solid green and purple line respectively. Their corresponding designs with maximal wavefunction overlaps are labeled by green and purple dot (b) Dependence of the wavelength and wavefunction overlap on the width of InAs QW at a fixed 2.3 nm wide $Ga_{0.6}In_{0.4}Sb$ QW. Design A is highlighted by blue and red dots (c) Dependence of the wavelength and wavefunction overlap on the width of $Ga_{0.6}In_{0.4}Sb$ QW at fixed 2.65 nm InAs QW thickness. Design A is highlighted by blue and red dots.

In contrast to W-shaped QWs, V-shaped QWs have four degrees of freedom in terms of design, instead of two: the thicknesses of InAs, $Ga_{0.6}In_{0.4}Sb$, $InAs_xSb_{1-x}$ and arsenic fraction ratio $x$.

Maximal wavefunction overlaps of triple V-shaped QW designs (**AlSb**/*InAs*/$Ga_{0.6}In_{0.4}$Sb/*$InAs_xSb_{1-x}$*/$Ga_{0.6}In_{0.4}$Sb/*InAs*/**AlSb**), for the wavelengths 6.2 μm, 7.5 μm (design B) and 9.0 μm are listed in the Table 1. The maximal wavefunction overlaps at 6.2 μm, 7.5 μm and 9.0 μm are 0.681, 0.642 and 0.608, respectively.

**Table 1.**

Layer structure of V-shaped QW composed of **AlSb**/*InAs*/$Ga_{0.6}In_{0.4}$Sb/*$InAs_xSb_{1-x}$*/$Ga_{0.6}In_{0.4}$Sb/*InAs*/**AlSb**, at wavelengths of 6.2 μm, 7.5 μm and 9.0 μm.

| λ | InAs | $Ga_{0.6}In_{0.4}Sb$ | $InAs_xSb_{1-x}$ | $x$ |
|---|---|---|---|---|
| 6.2 μm | 2.2 nm | 1.6 nm | 2.8 nm | 0.85 |
| 7.5 μm | 2.4 nm | 1.8 nm | 3.1 nm | 0.85 |
| 9.0 μm | 2.6 nm | 1.8 nm | 3.4 nm | 0.90 |

Therefore, the V-shaped design provides approximately 19 % higher wavefunction overlap than the W-shaped design; in particular 18.8 % (6.2 μm), 18.9 % (7.5 μm) and 19.2 % (9.0 μm). Due to the high dimensionality of the design space of the V-shaped design, we focus on the optimization of wavefunction overlap for the wavelength of 7.5 μm. The optimization procedure for the wavelengths of 6.2 μm and 9.0 μm is carried out analogically. Figure 3 (a) depicts the wavefunction overlap as a function of the InAs and $Ga_{0.6}In_{0.4}Sb$ QW widths at a fixed width of the middle $InAs_{0.85}Sb_{0.15}$ QW with a thickness of 3.1 nm. The designs corresponding to wavelength about 7.5 μm are labeled by the blue contour. The wavefunction overlap in the wavelength range 7.4-7.6 μm varies from 0.518 to 0.642. The design with the highest wavefunction overlap (design B) is highlighted with a gray dot on the blue curve. The maximum wavefunction overlap is comparable to the state-of-the-art recombination region of 3.5 μm ICL, hence the optimization compensates for the reduction introduced by the longer wavelength. Figures 3 (b) and 3 (c) present the horizontal and vertical cuts of Figure 3 (a) across Design B, respectively. In Figure 3 (b), wavelength quadratically increases with InAs thickness as the effective bandgap reduces. The wavefunction overlap on the other hand is a non-monotonic function with a maximum at around an InAs QW width of 1.5 nm. At smaller widths, their ground state energies become higher compared to the ground state energy of the middle InAsSb well. Hence, they are effectively decoupled and the total ground state is more confined in the middle InAsSb well rather than in the InAs wells. The wavefunction overlap slightly increases with the InAs width as the electron confinement redistributes from the middle InAsSb to the side wells, until the maximum value at 1.5 nm. Then for wider InAs QWs, it decreases as the electron confinement in InAs gets higher. Analogously, in Figure 3 (c), at fixed width of InAs QWs, the wavelength parabolically increases with $Ga_{0.6}In_{0.4}Sb$ QWs width, however it reaches its maximum at 3 nm, as the shift of the hole eigenstate in $Ga_{0.6}In_{0.4}Sb$ becomes negligible. Unlike the previous cut, the wavefunction overlap decreases monotonously with width of $Ga_{0.6}In_{0.4}Sb$ QWs as the electron wavefunctions become more decoupled. Unlike for W-shaped QW, variation of widths of electron and hole $Ga_{0.6}In_{0.4}Sb$ wells comparably contribute to change of the wavefunction overlap.

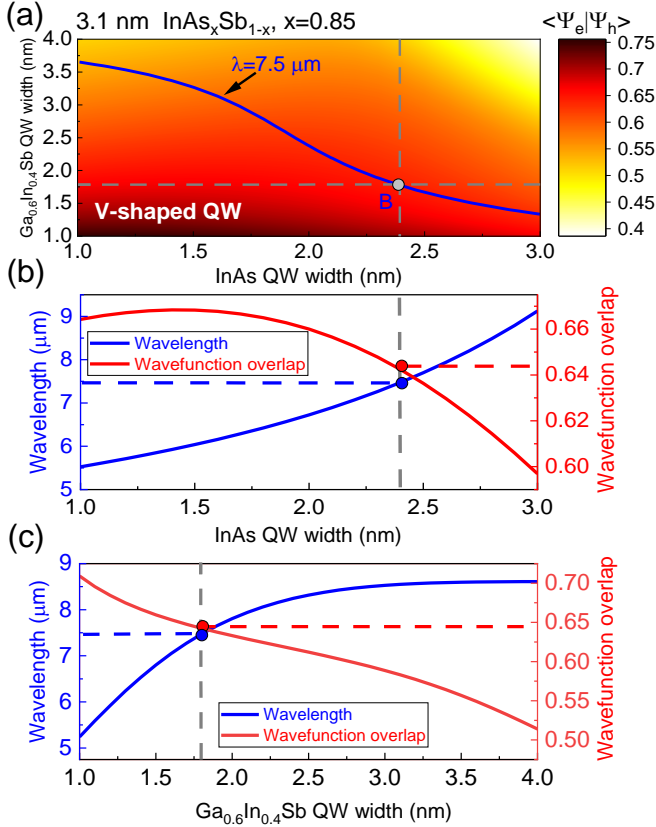

Fig. 3. (a) Dependence of the wavefunction overlap in a triple V-shaped QW for a variation of the electron InAs and hole $Ga_{0.6}In_{0.4}Sb$ QW widths at fixed 3.1 nm width of InAsSb electron QW and its arsenic fraction ratio of 0.85. The region corresponding to wavelength of 7.5 μm is marked with the solid, blue contour line. Design B, which corresponds to the highest wave-function overlap, is highlighted by a gray dot (b) Dependence of the wavelength and wavefunction overlap on width of InAs QWs at fixed 1.8 nm wide $Ga_{0.6}In_{0.4}Sb$ QWs. Design B is highlighted by blue and red dots (c) Dependence of the wavelength and wavefunction overlap on width of $Ga_{0.6}In_{0.4}Sb$ QWs at fixed 2.4 nm InAs QWs thickness. Design B is highlighted by blue and red dots.

Conversely, in Figure 4 (a) we keep the width of InAs electron wells and $Ga_{0.6}In_{0.4}Sb$ hole wells of Design B and vary the width of the middle InAsSb electron QW and its arsenic fraction ratio. Figures 4 (b) and 4 (c) display the horizontal and vertical cuts of Figure 4 (a) across Design B respectively.

Figure 4 (b) shows that both wavelength and wavefunction overlap almost parabolically change with the increase of the arsenic fraction ratio in the middle InAsSb QW. At arsenic fraction ratio of $x < 0.12$, the valence band edge is energetically higher in $InAs_xSb_{1-x}$ than in $Ga_{0.6}In_{0.4}Sb$. Hence, $InAs_xSb_{1-x}$ and two $Ga_{0.6}In_{0.4}Sb$ layers form one joint hole QW and the hole distribution is highest in $InAs_xSb_{1-x}$. In addition, the electron ground state in the middle well is significantly higher than in the side electron InAs wells, so only states from the side states are coupled. Both effects result in a low wavefunction overlap. At $x = 0.12$, $InAs_xSb_{1-x}$ switches to a hole barrier and the holes redistribute from the middle to the $Ga_{0.6}In_{0.4}Sb$ layers. For values $0.12 < x < 0.4$ the electron state in $InAs_xSb_{1-x}$ becomes lower and the middle state couples with the states from the InAs wells. Additionally, the valence band offset between $Ga_{0.6}In_{0.4}Sb$ and $InAs_xSb_{1-x}$ increases, which decouples hole wavefunctions more, until the wavefunction overlap maximum at $x \approx 0.4$. Larger arsenic content makes the electron $InAs_xSb_{1-x}$ QW deeper, leading to higher electron confinement in the middle well. For $x > 0.4$, the wavefunction overlap reduces since the middle electron well starts to decouple from the side wells. Regarding wavelength, for $x < 0.12$ the hole eigenstate is high. With an increase in arsenic content, the joint hole well switches to two significantly thinner $Ga_{0.6}In_{0.4}Sb$ wells, which lowers the ground states.

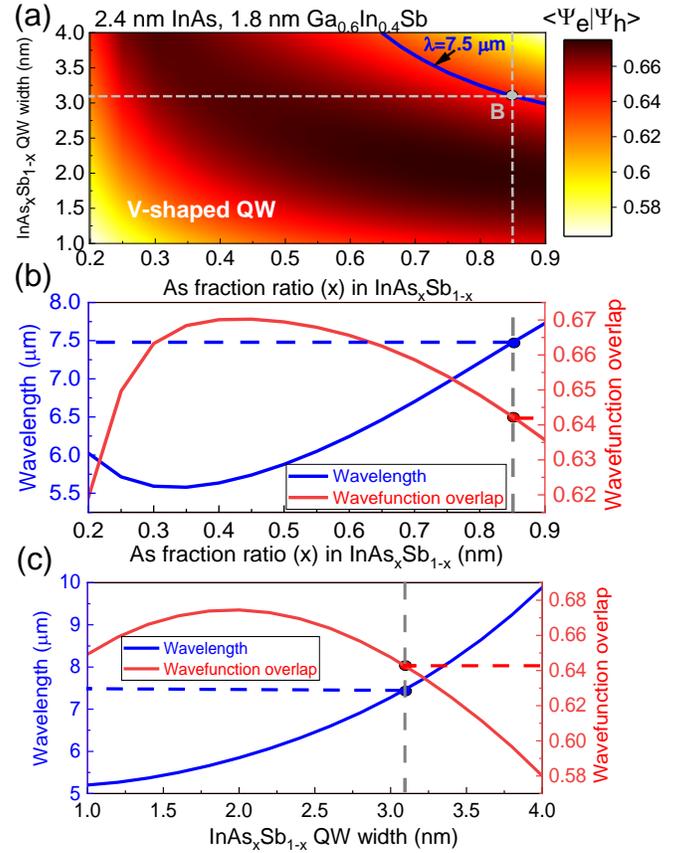

Fig. 4. (a) Dependence of the wavefunction overlap in a triple V-shaped QW for a variation of the electron $InAs_xSb_{1-x}$ QW width and its arsenic fraction ratio (x) at fixed width of 2.4 nm of InAs electron QW and 1.8 nm $Ga_{0.6}In_{0.4}Sb$ hole QW. The region corresponding to wavelength of 7.5 μm is marked with the solid, blue contour. The design with the highest wavefunction overlap is marked with the gray dot and the dashed line (design B) (b) Dependence of the wavelength and wavefunction overlap on arsenic fraction ratio (x) in 3.1 nm $InAs_xSb_{1-x}$ QW at fixed width of 2.4 nm of InAs electron QWs and 1.8 nm $Ga_{0.6}In_{0.4}Sb$ hole QWs. Design B is labeled with the blue and red dot (c) Dependence of the wavelength and wavefunction overlap on thickness of $InAs_{0.85}Sb_{0.15}$ at fixed width of 2.4 nm of InAs electron QWs and 1.8 nm $Ga_{0.6}In_{0.4}Sb$ hole QWs. Design B is labeled with blue and red dot.

This increases the transition energy (decreases wavelength). For $0.12 < x < 0.35$, the wavelength continues to decrease. The reason for this is that the downward shift of the electron state, caused by the shift of the conduction band edge of $InAs_xSb_{1-x}$, is still smaller than the shift of the hole states in $Ga_{0.6}In_{0.4}Sb$ wells, induced by raise of the middle $InAs_xSb_{1-x}$ barrier. At $x = 0.35$, the wavelength reaches its minimum. For $x > 0.35$ the shift of electron energy state surpasses the shift of hole states. Therefore, the transition energy starts decreasing (the wavelength increases). It should be noted that growth of InAsSb layers of a good crystal quality would be challenging at low arsenic fraction ratios as it would involve a significant compressive strain. In Figure 4 (c), the wavelength increases with the width of $InAs_{0.85}Sb_{0.15}$ QW as the effective band gap reduces. For $InAs_{0.85}Sb_{0.15}$ QW shorter than 2 nm, its energy level is significantly higher than those in side InAs wells. This decouples side electron wells and reduces the wavefunction overlap. With wider $InAs_{0.85}Sb_{0.15}$, its energy level moves downwards and couples with the states from InAs wells, reaching the maximum wavefunction overlap at width of $d_{InAsSb} \approx$ 2 nm. With increasing its width further, the wavefunction overlap starts decreasing because the electron probability in the $Ga_{0.6}In_{0.4}Sb$ QWs becomes smaller.

In addition to wavefunction overlap, the VISA was calculated for the optimized designs of both W-shaped and V-shaped structures by QuantCAD's software CADtronics [32].

Figure 5 presents the VISAs spectra of Designs A and B. In the 6-9 μm range, the V-shaped QW has a lower VISA than the W-shaped design. The highest difference is achieved at wavelength of 7.05 μm. At 6.2 μm, the VISA for W-shaped and V-shaped is approximately equal.

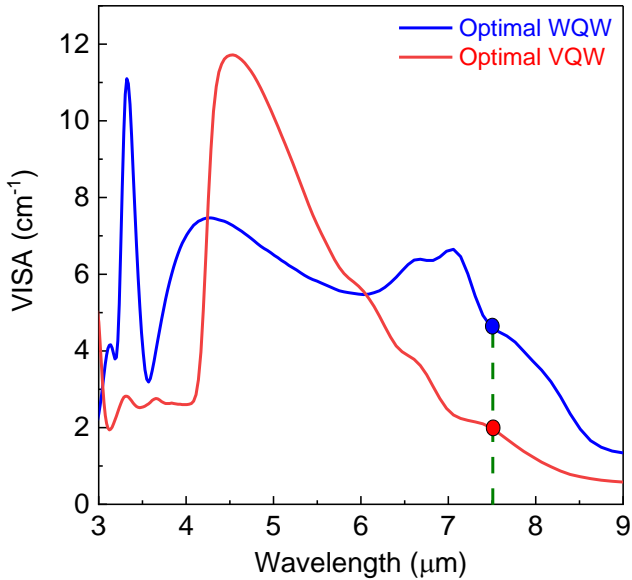

Fig 5. VISA spectra of W-shaped and triple V-shaped QW structures, optimized in terms of the wavefunction overlap. The VISA of V-shaped QW is significantly smaller at 7.5 μm.

However, at 7.5 μm, the absorption of the triple V-shaped QW ($\alpha_{VISA} = 2.0$ cm$^{-1}$) is significantly smaller than of the W-shaped QW ($\alpha_{VISA} = 4.65$ cm$^{-1}$).

## 3. Optimization of V-QW at applied electric field

To investigate the properties of the W-shaped and the triple V-shaped QWs as the active region of ICLs, wavefunction overlap and absorption are investigated under electric fields. To address the Stark shift, slightly asymmetrical W-shaped and V-shaped wells are considered as typically employed in ICL active region designs. The layer structure of the W-shaped QW is: **AlSb**/*InAs*/$Ga_{0.6}In_{0.4}Sb$/*InAs*/**AlSb**, with respective thicknesses of (**2.5**/*2.70*/*2.30*/*2.60*/**1.5**) nm, whereas the layer structure of triple V-shaped QW is **AlSb**/*InAs*/$Ga_{0.6}In_{0.4}Sb$/*InAs$_{0.85}$Sb$_{0.15}$*/$Ga_{0.6}In_{0.4}Sb$/*InAs*/**AlSb**, with thicknesses (**2.5**/*2.4*/*1.8*/*3.1*/*2.0*/*2.4*/**1.5**) nm.

The wavefunction overlaps for both symmetrical (s-) and asymmetrical (a-) W-shaped and triple V-shaped QW designs are listed in Table 2.

**Table 2**.

Wavefunctions overlaps of W-shaped and triple V-shaped QWs for symmetrical (s) and asymmetrical (a) designs. Under the electric field of 37 kV/cm, the two GaInSb hole QWs of V-shaped QW couple and the wavefunction overlap is maximal.

| Field ($\frac{kV}{cm}$) | s-WQW | a-WQW | s-VQW | a-VQW |
|---|---|---|---|---|
| 0 | 0.540 | 0.538 | 0.642 | 0.438 |
| 37 | | | | 0.605 |
| 50 | 0.538 | 0.539 | 0.422 | 0.437 |
| 100 | 0.532 | 0.537 | 0.385 | 0.393 |

The wavefunction overlap in the W-shaped QW slightly drops at higher electric field. However, V-shaped QW are substantially more sensitive to the electric field. This is a consequence of coupling between the hole ground states of $Ga_{0.6}In_{0.4}Sb$ wells. For the symmetrical structure, the hole states are coupled at zero field, and the overlap is maximal (0.642). For the asymmetrical structure, the states are coupled at field intensity of $F = 37$ kV/cm, and the maximal wavefunction overlap is 0.605. For higher or lower fields, the states are decoupled and one peak of the coupled hole wavefunction is effectively lost, which results in a drastic drop of the overlap. These changes in band structure are shown in Figure 6. However, it is worth pointing out that this coupling sweet spot has a certain window due to energy dispersion and thermal broadening.

Figure 7 shows absorption spectra of slightly asymmetrical W-shaped and triple V-shaped QWs for different electric fields. At the wavelength of 7.5 μm, the electric field barely affects the VISA. The VISA in W-shaped WQW is $\alpha_{VISA} = 4.50$ cm$^{-1}$. For the V-shaped QW it is $\alpha_{VISA} = 2.65$ cm$^{-1}$, implying that the surpression of VISA introduced by V-shaped design holds also under applied field.

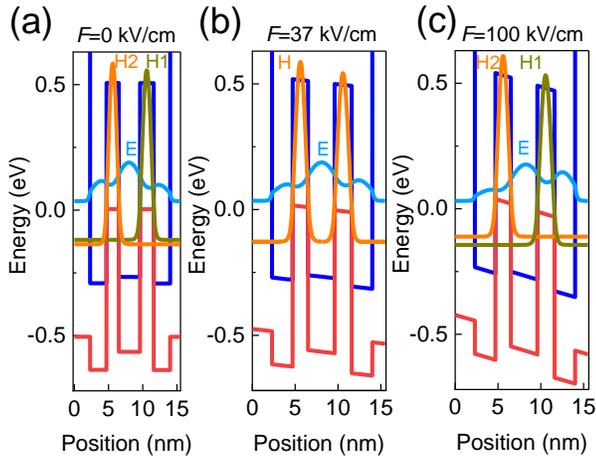

Fig. 6. Band structure of asymmetrical V-shaped QW at different applied electric fields: 0 (a), 37 kV/cm (b) and 100 kV/cm (c). At an electric field of 37 kV/cm, the hole states of the $Ga_{0.6}In_{0.4}Sb$ wells are coupled.

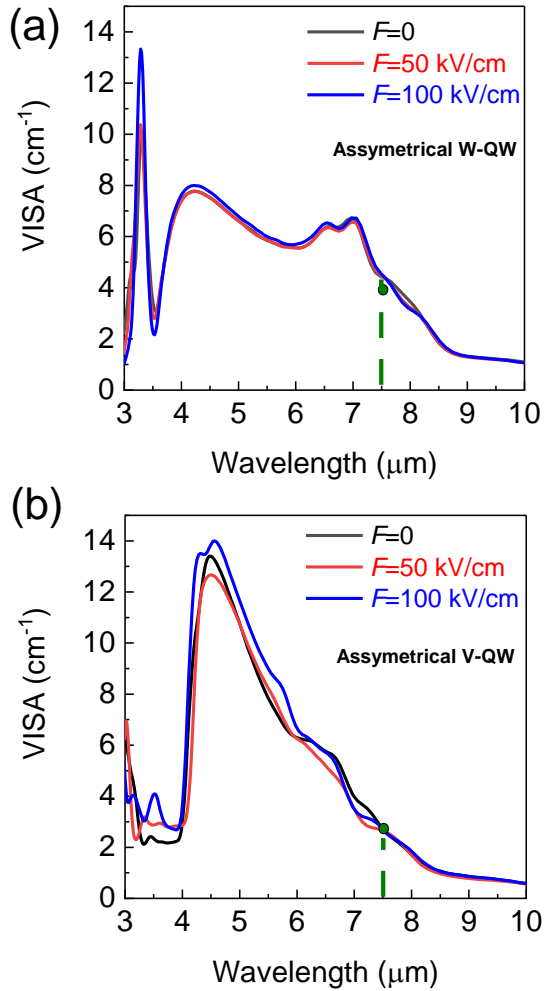

Fig. 7. VISA spectra of (a) W-shaped QW and (b) triple V-shaped QW design at different applied electric fields: 0, 50 and 100 kV/cm.

## 4. Conclusion

In summary, we have introduced a triple V-shaped QW design, **AlSb/*InAs*/$Ga_{0.6}In_{0.4}Sb$/*$InAs_xSb_{1-x}$*/$Ga_{0.6}In_{0.4}Sb$/*InAs*/AlSb**, for wavelength range 6-9 μm and compared their wavefunction overlap and VISA to the respective values for commonly used W-shaped well **AlSb/*InAs*/$Ga_{0.6}In_{0.4}Sb$/*InAs*/AlSb**. It was shown that V-shaped design provides approximately 19 % higher wavefunction overlap. In addition, 7.5 μm V-shaped design features 57 % lower VISA in than W-shaped QW. Specifically at 7.5 μm, the optimization of the design parameters is demonstrated to maximize the wavefunction overlap. Under applied electric field, a high sensitivity of wavefunction overlap on the electric field value is shown. As reported in Ref. 28, due to this reason, implementation of such V-shaped QW in the active region of an ICL can be challenging. However, this challenge could be addressed by adjusting the well widths if threshold voltage can be successfully predicted.

We are grateful to European Union's Horizon 2020 research and innovation program under the Marie Skłodowska-Curie grant agreement (no 956548, QUANTIMONY) for financial support.

**Declaration of competing interest**

The authors declare that they have no known competing financial interests or personal relationships that could have appeared to influence the work reported in this paper.

**References**

[1] R. Q. Yang, "Infrared laser based on intersubband transitions in quantum wells," *Superlattices Microstruct.* **17** (1), 77–83 (1995).

[2] J. Faist, F. Capasso, D. L. Sivco, C. Sirtori, A. L. Hutchinson, and A. Y. Cho, "Quantum Cascade Laser," *Science* **264**, 553 (1994).

[3] J. R. Meyer, W. W. Bewley, C. L. Canedy, C. S. Kim, M. Kim, C. D. Merritt, and I. Vurgaftman, "The interband cascade laser," Photonics **7** (3), 75 (2020).

[4] G. Wysocki, Y. Bakhirkin, S. So, F. K. Tittel, C. J. Hill, R. Q. Yang, and M. P. Fraser, "Interband cascade laser-based trace-gas sensor for environmental monitoring," *Appl. Opt.* **46**, 8202 (2007).

[5] M. von Edlinger, J. Scheuermann, R. Weih, L. Nähle, M. Fischer, J. Koeth, S. Höfling, and M. Kamp, "Interband cascade lasers for applications in process control and environmental monitoring," in *Light, Energy and the Environment Congress* (2015), *Paper No. EM2A.5.*

[6] B. Henderson, A. Khodabakhsh, M. Metsälä, I. Ventrillard, F. M. Schmidt, D. Romanini, G. A. D. Ritchie, S. L. Hekkert, R. Briot, T. Risby et al., "Laser spectroscopy for breath analysis:

[7]M. Kim, C. L. Canedy, W. W. Bewley, C. S. Kim, J. R. Lindle, J. Abell, I. Vurgaftman, and J. R. Meyer, "Interband cascade laser emitting at λ=3.75 μm in continuous wave above room temperature," *Appl. Phys. Lett*. **92**, 191110 (2008).

[8]W. W. Bewley, C. L. Canedy, C. S. Kim, M. Kim, C. D. Merritt, J. Abell, I. Vurgaftman, and J. R. Meyer, "High-power room-temperature continuous-wave mid-infrared interband cascade lasers," *Proc. SPIE* 8374, 83740H (2012).

[9]I. Vurgaftman, W. W. Bewley, C. L. Canedy, C. S. Kim, M. Kim, C. D. Merritt, J. Abell, J. R. Lindle and J. R. Meyer, "Rebalancing of internally generated carriers for mid-infrared interband cascade lasers with very low power consumption" *Nat. Commun*. **2**, 585 (2011).

[10]R. Weih, M. Kamp, and S. Höfling, "Interband cascade lasers with room temperature threshold current densities below 100 A/cm$^2$," *Appl. Phys. Lett*. **102**, 231123 (2013).

[11]H. Knötig, J. Nauschütz, N. Opačak, S. Höfling, J. Koeth, R. Weih, and B. Schwarz, "Mitigating valence intersubband absorption in interband cascade lasers," *Laser Photonics Rev*. **16** (9), 2200156 (2022).

[12]A. Windischhofer, N. Opačak, B. Schwarz, "Charge Transport in Interband Cascade Lasers: An Ab-Initio Self-Consistent Model", *Laser & Photonics Rev*. **19**, 3, 2400866 (2024).

[13]J. R. Meyer, C. A. Hoffman, F. J. Bartoli, L. R. Ram-Mohan, "Type-II quantum-well lasers for the mid-wavelength infrared" *Appl. Phys. Lett*. **67**, 757-759 (1995).

[14]J. Nauschütz, H. Knötig, R. Weih, J. Scheuermann, J. Koeth, S. Höfling, and B. Schwarz, "Pushing the room temperature continuous-wave operation limit of GaSb-based interband cascade lasers beyond 6 μm," *Laser Photonics Rev*. **17** (4), 2200587 (2022).

[15]A. Bader, L. Steinbrecher, F. Rothmayr, Y. Rawal, F. Hartmann, A. Pfenning, and S. Höfling, *III-V semiconductor mid-infrared interband cascade light emitters and detectors, in Infrared Remote Sensing and Instrumentation XXIX,* **11830**, pp.113 – 122, International Society for Optics and Photonics, SPIE (2021).

[16]M. Dallner, F. Hau. S Höfling and M. Kamp, "InAs-based interband-cascade-lasers emitting around 7 μm with threshold current densities below 1 kA/cm2 at room temperature" *Appl. Phys. Lett*. **106**, 041108 (2015).

[17]Y. Shen, R. Q. Yang, J. D. Steward, S. D. Hawkins and A. J. Muhowski, "Room temperature interband cascade lasers near 7.7 μm and dependence on structural quality," *Semicond. Sci. Technol*. **40**, 055002 (2025).

[18]J. A. Massengale, Y, Shen, R. Q. Yang, S. D. Hawkins, J. F. Klem, "Long wavelength interband cascade lasers ", *Appl. Phys. Lett*. **120**, 091105 (2022).

[19]Y. Shen, J. A. Massengale, R. Q. Yang, S. D. Hawkins, and A. J. Muhowski, "Pushing the performance limits of long wavelength inteband cascade lasers using innovative quantum well active regions," *Appl. Phys. Lett*. **123**, 041108 (2023).

[20]R. Q. Yang, L. Li, W. Huang, S. M. Shazzad Rassel, J. A. Gupta, A. Bezinger, X. Wu, S. Ghasem Razavipour, and G. C. Aers, "InAs-based interband cascade lasers," *IEEE J. Sel. Top. Quantum Electron*. **25**, 1200108 (2019).

[21]E. Tournié, L. M. Bartolome, M. Rio Calvo, Z. Loghmari, D. A. Diaz-Thomas, R. Teissier, A. N. Baranov, L. Cerutti and J-B. Rodriguez, "Mid-infrared III-V semiconductor lasers epitaxially grown on Si substrates," *Light Sci. Appl*. **11**(1):165 (2022).

[22]M. Fagot, D. A. Diaz-Thomas, A. Gilbert, G. Kombila, M. Ramonda, Y. Rouillard, A. N. Baranov, J-B. Rodriguez, E. Tournié and L. Cerutti, "Interband cascade lasers grown simultaneously on GaSb, GaAs and Si substrates," *Opt. Express* **32**(7), 11057-11064 (2024).

[23]A. Spott, J. Peters, M. L Davenport, E. J. Stanton, C. D. Merritt, W. W. Bewley, I. Vurgaftman, C. S. Kim, J. R. Meyer, J. Kirch, L. J. Mawst, D. Botez and J. E. Bowers, "Quantum cascade laser on silicon," *Optica* **3**(5), 545-551 (2016).

[24]Z. Loghmari, J-B. Rodriguez, A. N. Baranov, M. Rio-Calvo, L. Cerutti, A. Meguekam, M. Bahriz, R. Teissier and E. Tournié, "InAs-based quantum cascade lasers grown on on-axis (001) silicon substrate," *APL Photonics* **5**, 041302 (2020).

[25]J. R. Meyer, C. A. Hoffman, F. J. Bartoli, L. R. Ram-Mohan, "Type-II quantum-well lasers for the mid-wavelength infrared," *Appl. Phys. Lett*. **67**, 757–759 (1995).

[26]I. Vurgaftman, J. R. Meyer, L. R. Ram-Mohan, "Mid-IR Vertical-Cavity Surface-Emitting Lasers," *IEEE J. Quantum Electron*. **34**, 147–156 (1998).

[27]Y. Jiang, L. Li, Z. Tian, H. Ye, L. Zhao, R. Q. Yang, T. D. Mishima, M. B. Santos, M. B. Johnson and K. Mansour, "Electrically widely tunable interband cascade lasers," *Journ. Appl. Phys*. **115**, 113101 (2014).

[28]M. Motyka, K. Ryczko, M. Dyksik, G. Sęk, J. Misiewicz, R. Weih, M. Dallner, S. Höfling, M. Kamp, "On the modified active region design of interband cascade lasers" *J. Appl. Phys*. **117**, 084312 (2015).

[29]M. Dyksik, "Electrical tuning of the oscillator strength in type II InAs/GaInSb quantum wells for active region of passively mode-locked interband cascade lasers", *Jpn. J. Appl. Phys*. **56**, 110301 (2017).

[30]S. Birner, T. Zibold, T. Andlauer, T. Kubis, M., A. Trellakis, and P. Vogl, "nextnano: General Purpose 3-D Simulations", *IEEE Trans. Electron. Devices* **54**, 2137 (2007).

[31]I.Vurgaftman and J. R. Meyer, "Band parameters for III-V compound semiconductors and their alloys," *J. Appl. Phys*. **89**, 11 (2001).

[32]J. T. Olesberg, M. E. Flatté, T. C. Hasenberg and C. H. Grein, "Mid-infrared InAs/GaInSb separate confinement heterostructure laser diode structures," *Journ. Appl. Phys*. **89**(6), 3283-3289 (2001).